\documentclass{imammb}
\jno{dqnxxx}
\usepackage{natbib}
\usepackage{epsfig}

\usepackage{amsmath,amsthm,bm,mathrsfs}

\numberwithin{equation}{section}

\begin{document}

\title{Combined therapies of antithrombotics and antioxidants delay \emph{in silico} brain tumor progression}

\author{ {\sc Alicia Mart\'{\i}nez-Gonz\'{a}lez$^*$}\\[2pt]
Departamento de Matem\'aticas, Universidad de Castilla-La Mancha \\
ETSI Industriales, Avda. Camilo Jos\'e Cela 3, 13071 Ciudad Real, Spain.\\
$^*${\rm Corresponding author: alicia.martinez@uclm.es} \\[6pt]
 {\sc Mario Dur\'{a}n-Prado}\\[2pt]
Dpto. de Ciencias M\'{e}dicas, Facultad de Medicina, Universidad de Castilla-La Mancha, 13071 Ciudad Real, Spain.\\[6pt]
{\sc Gabriel F. Calvo}\\[2pt]
Departamento de Matem\'aticas, Universidad de Castilla-La Mancha \\
ETSI Caminos, Canales y Puertos, Avda. Camilo Jos\'e Cela 3, 13071 Ciudad Real, Spain.\\[6pt]
{\sc Francisco J. Alca\'{\i}n} \\[2pt]
Dpto. de Ciencias M\'{e}dicas, Facultad de Medicina, Universidad de Castilla-La Mancha, 13071 Ciudad Real, Spain. \\[6pt]
{\sc Luis A. P\'erez-Romasanta} \\[2pt]
Servicio de Oncolog\'ia Radioter\'apica, Hospital General Universitario de Salamanca, Spain \\[6pt]
{\sc V\'{\i}ctor M. P\'{e}rez-Garc\'{\i}a}\\[2pt]
Departamento de Matem\'aticas, Universidad de Castilla-La Mancha \\
ETSI Industriales, Avda. Camilo Jos\'e Cela 3, 13071 Ciudad Real, Spain.\\
{\rm [Received on  XXXXX]} \vspace*{6pt}}
\pagestyle{headings}
\markboth{ A. MART\'INEZ-GONZ\'{A}LEZ \emph{et al.}}{\rm COMBINED THERAPIES DELAY BRAIN TUMOR PROGRESSION}

\maketitle

\begin{abstract}
{Glioblastoma multiforme, the most frequent type of primary brain tumor, is a rapidly evolving and spatially heterogeneous high-grade astrocytoma that presents areas of necrosis, hypercellularity and microvascular hyperplasia. The aberrant vasculature leads to hypoxic areas and results in an increase of the oxidative stress selecting for more invasive tumor cell phenotypes.  In our study we assay {\em in silico} different therapeutic approaches which combine antithrombotics, antioxidants and standard radiotherapy. To do so, we have developed a biocomputational model of glioblastoma multiforme  that incorporates the spatio-temporal interplay among two glioma cell phenotypes corresponding to oxygenated and hypoxic cells, a necrotic core and the local vasculature whose response evolves with tumor progression. Our numerical simulations predict that suitable combinations of antithrombotics and antioxidants may diminish, in a synergetic way, oxidative stress and the subsequent hypoxic response. This novel therapeutical strategy, with potentially low or no toxicity, might reduce tumor invasion and further sensitize glioblastoma multiforme to conventional radiotherapy or other cytotoxic agents, hopefully increasing median patient overall survival time.}
Glioblastoma multiforme, antithrombotic, radiotherapy, antioxidants, combined therapies
\end{abstract}

\section{Introduction}
Glioblastoma Multiforme (GBM), a World Health Organization (WHO) grade IV astrocytic glioma, is the most aggressive and frequent primary brain tumor \citep{WHO}.  GBM may develop (spanning from 1 year to more than 10 years) from lower-grade astrocytomas (WHO grade II) or anaplastic astrocytomas (WHO grade III), but more frequently, it manifests de novo. The median overall survival ranges from 12 to 15 months after diagnosis despite using the current standard care. It includes maximal safe resection followed by radiotherapy in combination with the chemoterapeutic (alkylating) agent temozolomide \citep{Meir}. After treatment, relapse occurs typically within a few months due to the GBM radio and chemoresistance and its remarkable infiltrative nature.
\par
At the tissue level, the presence of necrosis and microvascular hyperplasia is characteristic of GBM, in contrast to lower-grade gliomas where peritumoral vascular damage is infrequent or absent. Histological samples of GBM frequently show thrombosed vessels within necrotic cores.
Tumor cells actively migrate away from oxygen-deficient (hypoxic) regions, originated after vascular injury. They form hypercellular regions that surround the necrotic cores \citep{pseudopalisading2}.

 \par
At the molecular level, adaptation of tumor cell subpopulations to strong spatio-temporal variations of oxygen availability is mediated by the hypoxia-inducible factor (HIF-1). This complex activates the transcription of hundreds of genes that play key roles, among others, in cell death avoidance, genetic instability, vascularization, glucose metabolism, pH regulation, immune evasion and invasion \citep{Kelly,Semenza2012}.  Under acute hypoxia, tumor cells will tend to cease or reduce their proliferation rate as a means to decrease oxygen consumption. This mechanism relies primarily on the subunit HIF-1$\alpha$ of the heterodimer HIF-1, which arrests DNA replication in the presence of oxygen stress \citep{Hubbi2013}. Moreover, the levels of free radicals increase during cycles of hypoxia and there is evidence that they are also molecular switches influencing the stabilization of  HIF-1$\alpha$ \citep{Wilson}. 
\par 
In contrast to normal cells, the survival of GBM cells is favored by an increase in free radicals since the oxidative stress forces the cells to produce antioxidant enzymes  such as catalase and superoxide dismutase to control the exacerbated level of free radicals \citep{Kovacic,Schumacker}. The increase in these antioxidant enzymes strongly interferes with the action of radio/chemotherapy, that kill tumor cells by inducing an increase in free radicals  \citep{Tennant2010}. In addition, free radicals inactivate the tumor suppressor protein p53, enabling tumor cells  to escape apoptosis  \citep{Cobbs}. Thus, disrupting this signaling process in tumors, may be expected to promote hypoxia-induced death \citep{BCR} and to restore redox homeostasis, turning off the hypoxic response. 
\par
 Many evidences indicate that the invasive ability of GBM  and its resistance to chemo and radiotherapy is mostly due to phenotypic changes that are intimately linked to the presence of  hypoxia, initiated by oxygen deprivation after a vaso-occlusion event. Under these conditions, the levels of HIF-1$\alpha$ increase driving the expression of many pro-angiogenic factors such as the Vascular Endothelial Growth Factor (VEGF). The accumulation of VEGF promotes an aberrant neovascularization  \citep{Semenza} composed of a high density of non functional microvessels which are leaky and devoid of pericytes. 
 This structure allows the contact between blood and tumor cells and initiates the coagulation cascade and the formation of thrombi, a major cause of patient death \citep{Jenkins,Young}. 
Furthermore, the transmembrane protein tissue factor (TF) not only activates the extrinsic pro-coagulant pathway but also contributes to GBM cell invasion and neovascularization \citep{Mittelbronn}.
 In general, thromboembolism, i.e. vaso-occlusion events of large vessels, is recognized as a major complication of cancer and a common cause of death in cancer patients. There is strong evidence linking venous thromboembolism (VTE) and malignancy \citep{Khorana,Green}. 
 \par 
Thromboembolic complications include a broad spectrum of clinical problems, a fact that has lead to the use of different ways of thromboprophylaxis \citep{Khorana,Green} 
for cancer patients. 
\citet{Bastida} proved that glioma cell lines secrete pro-thrombotic factors, and in fact,
glioma patients have a high incidence of VTE, with several studies suggesting that 25\%-30\% of these patients suffer thromboembolic events \citep{Streiff,Simanek}. 
\par
It is also known that the more tumoral tissue is removed during surgery of high-grade gliomas the less-likely are the patients to die from VTE \citep{Brose,Simanek}. Thus, all this body of evidence has led to the consideration of thromboprophylaxis using anticoagulants, such as for instance low molecular weight heparine (LMWH), for glioma patients, specially in those with a high potential risk of developing VTE \citep{Hamilton,Batchelor,Jenkins,Khorana}.
\par
In addition to their potential use to diminish the risk of VTE, LMWH exerts a direct effect on tumor cells and tumor stroma, as it can directly kill cancer cells and also inhibit the neovascularization process, which downregulates cell invasion \citep{Santos,Svensson,Green}. In this context, there have been a limited number of small-scale clinical studies evaluating the safety of antithrombotic therapy in GBM patients with a VTE risk supporting the safety of the approach \citep{Schneider}. However, no extra benefits beyond the substantial reduction in the VTE risk have been proven yet.
\par
In our previous study \citep{Alicia}, we developed a minimal mathematical model of GBM progression  incorporating the evolution of different tumor cell subpopulations and chemicals in the tumor microenvironment. The implications of that work are that antithrombotics would have a restrain effect on tumor progression, thus allowing a better local control of the tumor. However the potential gain achieved by the use of antithrombotics alone is modest. In this paper we will discuss a way to potentially obtain a much more substantial tumor control by targetting simultaneously the vessel occlusion events and the accumulation of hypoxia inducible factors.  

\par

Thus, in this paper we study {\em in silico} a more elaborated system, including two different cell phenotypes, the oxygen concentration in the brain tissue and a dynamic vasculature to assess  the potential effect of combined therapies targeting simultaneously the vessel-occlusion events and the ``normalization" therapy (antioxidation) together with standard radiotherapy as an example of cytotoxic therapy.
\par
\begin{figure*}[t]
\begin{center}
\epsfig{file=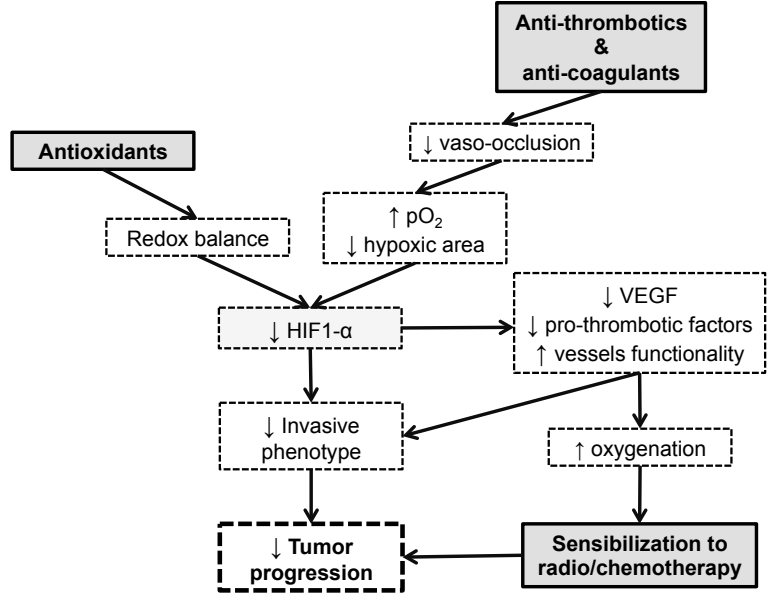,scale=0.4} 
\end{center}
\caption{{\bf Schematic description of how a combined treatment of antithrombotics and antioxidants may slow down GBM invasion and sensitize the tumor to radio and chemotherapy.} \label{fig:therapies}}
\end{figure*}
\par
Our conceptual framework is summarized in Fig. \ref{fig:therapies}. 
Vascular regularizing therapies and antioxidants follow different pathways aimed at reducing tumor hypoxia by means of directly impacting on HIF-1$\alpha$ levels. This  effect would decrease oxidative stress and the subsequent hypoxic response.  To achieve this goal, antioxidants act on the oxidation balance whereas antithrombotics reduce vaso-occlusions and hypoxic areas. The overall effect is an increment in oxygenation. Both therapies should ameliorate vessel functionality by reducing pro-thrombotic and angiogenic factors. In addition, the higher oxygen levels and decreased oxidative stress may provide the extra benefit of a sensitization to radio and chemotherapy, that might result in a synergistic response to the treatment.
\par
Our plan in this paper is as follows. First, in Sec. \ref{sec:themodel} we present the basic model equations describing tumor progression in detail and discuss the assumptions behind them. Next, in Sec. \ref{sec:thetherapies} we discuss how to incorporate the different therapies into the model. Parameter estimation is addressed in Sec. \ref{sec:theparameters}. 
Once the model is set up and realistic parameter ranges identified we move on to studying the dynamics; the main results being outlined in Sec. \ref{sec:theresults}. Finally, in Sec. \ref{sec:theconclusions} we discuss the implications of our findings and summarize our conclusions.


\section{Mathematical model of tumor progression}
\label{sec:themodel}
In our work, we propose a mathematical model that incorporates the spatio-temporal interplay among two tumor cell phenotypes corresponding to well oxygenated and hypoxic cells, a necrotic core, the oxygen distribution and a dynamic vasculature.

\begin{figure*}[t]
\begin{center}
\epsfig{file=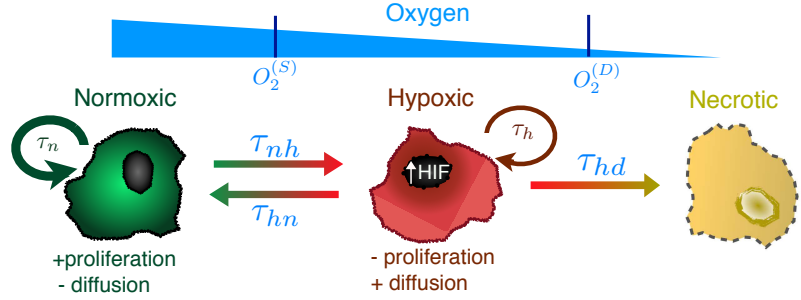,scale=0.35} 
\end{center}
\caption{{\bf Oxygen pressure influences the phenotype of a tumor cell.} Depending on the oxygen pressures various switching mechanisms arise coupling the populations:  Normoxic to hypoxic $S_{nh}$ for oxigen concentrations $O_2^{(S)}$ with characteristic time $\tau_{nh}$; hypoxic to normoxic $S_{hn}$ above $O_2^{(S)}$ with characteristic time $\tau_{hn}$ and hypoxic to necrotic $S_{hd}$ for pressures below $O_2^{(D)}$ with characteristic time $\tau_{hd}$. High oxygen pressures favor the existence of less mobile phenotypes with shorter doubling times $\tau_n$. On the contrary, cells respond to low oxygen pressures by expressing more motile phenotypes with larger doubling times $\tau_h$.\citep{pseudopalisading1}. \label{fig:presentation}}
\end{figure*}

\subsection{Cell phenotypes}
The so-called migration/proliferation dichotomy in tumor cells refers to the commonly assumed fact that cancer cells do not move and proliferate simultaneously \citep{Giese}. The switch between a more proliferative and a more invasive phenotype cannot be only mutation driven \citep{Hatzikirou,Onishi} and in fact it has been suggested  that invasive glioma cells are able to revert to a proliferative cellular program and vice versa, depending on the environmental stimuli \citep{Giese,Keunen}. Specifically, several studies have linked hypoxia to the invasive behavior of different types of tumors and their relationship with metastasis and negative prognosis \citep{Bristow,Kalliomaki, Elstner}. 
\par
Following these biological facts we will assume in our model that the heterogeneous oxygen distribution is the driving force that triggers phenotypic changes of tumor cells as shown schematically in Fig. \ref{fig:presentation}. Thus, the local oxygen concentration will be considered to have a major influence on the cell phenotype expression. It will be assumed to induce spatio-temporal  variations of phenotypic expressions via the HIF-1$\alpha$ activation/deactivation pathways. Therefore, will focus ourselves only on two dominant tumor cell phenotypes. A first phenotype with a spatio-temporal density described by a nonnegative function $C_n(x,t):{\mathbb{R}}^{2}\to{\mathbb{R}_{0}^{+}}$ corresponds to more proliferative tumor cells. A second phenotype will be described by (a nonnegative) function $C_h(x,t):{\mathbb{R}}^{2}\to{\mathbb{R}_{0}^{+}}$ accounting for tumor cells which are more mobile and resistant to therapies. Both phenotypes have described to have a major role in GBM progression  \citep{DeBerardinis,Giese,Keunen,Onishi}. For convenience, we will refer to them somewhat imprecisely as ``normoxic" and ``hypoxic" phenotypes since they correspond to the limit behavior of the tumor cell density after either very long oxygenation periods ($C_n(x,t)$) or persistent oxygen deprivation periods ($C_h(x,t)$).
 \par
The equations governing the interplay between the dominant phenotypes are
\begin{subequations}
\label{themodel}
\begin{eqnarray}
	\frac{\partial C_n}{\partial t}  & =  &  D_n \nabla^2 C_n + \frac{C_n}{\tau_n} \left(1 - \frac{C_n + C_h+C_d}{C^{\left(M\right)}}\right)  + \frac{S_{hn}}{\tau_{hn}}  C_h-  \frac{S_{nh}}{\tau_{nh}} C_n, \label{normox}  \\
	\frac{\partial C_h}{\partial t}  & =  & D_h \nabla^2 C_h +\frac{C_h }{\tau_h} \left(1 - \frac{C_n + C_h+C_d}{C^{\left(M\right)}}\right) -  \frac{S_{hn}}{\tau_{hn}}  C_h  + \frac{S_{nh}}{\tau_{nh}} C_n-  \frac{S_{hd}}{\tau_{hd}}C_h . \label{eqhipox}
	\end{eqnarray}
The first right-hand-side terms in Eqs. (\ref{normox}) and (\ref{eqhipox}) account for cellular motility. Since the hypoxic phenotype is more migratory than the normoxic one \citep{Berens,Giese,Bristow,Gorin}, the hypoxic cell diffusion coefficient $D_h$ should be chosen to be larger than the normoxic one $D_n$. This is a key assumption of our model whose implications will be elucidated in detail.
\par
The second terms in Eqs. (\ref{normox}) and (\ref{eqhipox}) are classical logistic growth terms for the tumor cell populations with proliferation times $\tau_n$ and $\tau_h$, respectively, and a maximum cell density $C^M$ that the brain tissue can accommodate. The migration-proliferation dichotomy suggests that $\tau_n < \tau_h$; this is also supported by the observation that hypoxic cells reduce their proliferation rate by arresting DNA replication \citep{Hubbi2013}. Since growth is assumed to be space-limited we incorporate also the necrotic tissue density $C_d(x,t)$ into the saturation terms (see below).
\par
The third and fourth terms in Eqs. (\ref{normox}) and (\ref{eqhipox}) represent the phenotypic switch under different oxygenations. 
The {\em switch} functions $S_{nh}$, $S_{hn}$ depend on the oxygen pressure. Under low oxygen conditions ($O_2^{(S)} < 7$ mmHg), normoxic cells change their phenotype to the hypoxic one, whereas above $O_2^{(S)}$ hypoxic cells recover their oxic phenotype.  $\tau_{nh}$ and $\tau_{hn}$ denote the characteristic switching times of these processes.  Usually $\tau_{nh}$ is much shorter than $\tau_{hn}$ since cells incorporate mechanisms to respond almost instantaneously to the absence of oxygen. However the reverse process is not so fast, specially in cells that have already suffered several oxygen deprivation episodes \citep{Hsieh}. \par
 
 To close the model we have to incorporate an equation for the necrotic tissue density $C_d(x,t)$
 \begin{equation}
\frac{\partial C_d}{\partial t}  =  \frac{S_{hd}}{\tau_{hd}}C_h .  \label{death} 
\end{equation}
\end{subequations}
Hypoxic cells (\ref{eqhipox}) feed the necrotic tissue in Eq. (\ref{death}) under persistent anoxic ($O_2^{(D)}<0.7$ mmHg) conditions. This is accounted for by the switch function $S_{hd}$ and a rate $1/\tau_{hd}$.  The explicit form of the switch functions will be taken to be \citep{Alicia}
\begin{subequations}
\begin{eqnarray}
 S _{nh} (O_2) & = & \frac{1}{2}\left[1-\tanh\left(\frac{O_2-O_2^{(S)}}{\Delta O_2}\right)\right], \\
  S _{hn} (O_2) & = & \frac{1}{2}\left[1+\tanh\left(\frac{O_2-O_2^{(S)}}{\Delta O_2}\right)\right], \\
  S _{hd} (O_2) & =  & \frac{1}{2}\left[1-\tanh\left(\frac{O_2-O_2^{(D)}}{\Delta O_2}\right)\right]
 \end{eqnarray}
 \end{subequations}
were the three step-like switch functions $S_{nh}$, $S_{hn}$ and $S_{hd}\in[0,1]$ and 
the parameter $ \Delta O_2$ described the characteristic range of oxygen variations where the transitions occur. 
 \par  

\subsection{Modelling oxygen distribution through the tissue}

To simplify the analysis, yet retaining the basic interplay of the key underlying biological processes in GBM,  we will stick to a one dimensional geometry with the oxygen flow coming from a network (1D lattice) of vessels. In the brain, spatio-temporal inhomogeneities arise in the parameter values (e.g. different propagation speeds in white and gray matter which also vary in time as the tumor progresses) and anisotropies (e.g. on the diffusion tensor with preferential propagation directions along white matter tracts \citep{Painter}). However, those complexities would add on top of the phenomena to be described in this paper. Thus, in what follows, we will focus our study on a representative spatial section which will provide key spatial and time metrics enabling us to incorporate, at a later stage, the action of various therapeutic modalities. 
\par
The spatio-temporal evolution of oxygen $O_2(x,t)$ is similar to the one chosen in \citet{Alicia}, although now oxygen sources are located on the blood vessels, and is governed by
\begin{equation}
	\frac{\partial O_2}{\partial t} =D_{O_2} \nabla^2 O_2 -  \frac{\alpha_n C_n + \alpha_h C_h}{O_2^{\left( T \right)} + O_2}O_2  +  J\chi^C (\chi^F O_2^{v_i}-O_2). \label{oxygen}
\end{equation}

The  first term in the right-hand-side of Eq. (\ref{oxygen}), accounts for oxygen diffusion in the brain tissue and assumes an homogenous and isotropic diffusion coefficient $D_{O_2}$. The second term models the oxygen consumption by both normoxic  and hypoxic cells at rates $\alpha_n$ and  $\alpha_h$, respectively. The saturation Michaelis-Menten constant $O_2^{(T)}$ corresponds to the oxygen pressure level at which the reaction rate is halved. The third term describes the oxygen flow from the vessels to the tissue.  The oxygen pressure in the $i^{th}$ blood vessel will be denoted as $O_2^{v_i}(t)$. 
  The multiplicative function $J(x)$  in Eq. \eqref{oxygen} accounts for the spatial distribution of the oxygen supply and depends on the blood vessels positions $p_i$ and their sizes $v_i$.  Here, $J$ will be taken to be a combination of Gaussian functions of the form
 \begin{equation}\label{J}
 J(x)=J_{O_2} \sum_{i=1}^Ne^{-(x-p_i)^2/v_i^2} ,
 \end{equation}
with $J_{O_2}=10^{-6}$ $s^{-1}$ as the estimation for the oxygen exchange coefficient which is considered to be constant in time and the same for all of the vessels.  The positions $p_i$ of the vessels will be considered to be equidistant, with separations of 300 $\mu$m. The vessel widths $v_i$ (being capillaries), will be taken to be 30 $\mu$m in diameter. 
Since oxygen diffuses along and through the vasculature it is reasonable to think that vessels of similar size will have similar nutrient concentrations. Therefore, $O_2^{v_i}$ is considered to be constant along time for all $i$.
Furthermore, $\chi^C$ and $\chi^F$ in Eq. (\ref{oxygen}) are factors accounting for
 the chronic and fluctuating hypoxia induced by the tumor and will be explained in depth in Section \ref{vascu}.
  \par
 Oxygen diffuses successively through the intracellular fluid, cell membranes and cytoplasm which have abrupt spatial variations. 
 However, previous works \citep{Pogue,Das,Powathil} have proved that using an average diffusion coefficient for oxygen as in Ec. \eqref{oxygen} provides a good approximation to the diffusion process.
 
 Given that mean oxygen pressure in arterial blood is around 95 mmHg \citep{Kimura} and venous values are around 30-40 mmHg, we will employ for the oxygen pressure within the capillaries a constant value  $O_2^{v_i} =$ 80 mmHg.
 
\subsection{Vasculature and coagulation}
\label{vascu}
The sources of nutrients and oxygen in the brain are the blood vessels (capillaries in here). We will also assume that oxygen concentration in the capillary network does not follow the variations induced by the cardiac pumping in the major blood vessels \citep{Das}, because of the fast time scale of those variations.
\par
Thrombi formation leading to the developing of pseudopalisading structures \citep{pseudopalisading1,pseudopalisading2} may be in part connected with the vascular remodeling induced by tumor cells. The pro-angiogenic stimuli give rise to hyperplasia in the endothelial cells which lose cell to cell unions causing vascular permeability and a malfunctioning of the blood-brain barrier (BBB). The generated orifices allow the blood to be in contact with the tumor cells. It results in plasma coagulation factors in direct connection with the tumor tissue. As a consequence, fibrin coagulates and platelets aggregate. 
The tumor secrets TF that activate the so-called extrinsic coagulation pathway what provides an additional source for the strong pro-thrombotic activity of GBM \citep{Ruf}.
\par
We will consider that vessel functionality decreases as tumor cell density increases. Once cells accumulate around the vessels the constitutive expression of TF by cancer cells will trigger local and systemic activation of the coagulation cascade as described above.
\par

In our approach, blood vessels will be first assumed to become semifunctional when the tumor cell density is above a certain threshold $C^{(F)}$ implying the onset of fluctuating hypoxia. The step-like switch function $\chi^F$ reproduces this phenomena depending on a cellular density $C^{(F)}$ and a random dynamic noise $z$ ($0<z<1$) given by the equation
\begin{subequations}
\label{chis}
 \begin{equation} 
  \chi^F(C_n+C_h+C_d,t^*)   =  1-z\frac{1}{2}\left[1+\tanh\left(\frac{(C_n+C_h+C_d)-C^{(F)}}{\Delta c}\right)\right] . \label{chi^F}
  \end{equation} 
We will also assume that when the tumor cell density around the vessel grows beyond a higher threshold value $C^{(C)}$, the coagulation process is irreversible and  chronic hypoxia sets in due to the complete lack of oxygen flow. This process is modelled by a step-like switch function $\chi^C$ given by 
\begin{equation}
   \chi^C(C_n+C_h+C_d,t^*)   =  \min \left\{ \frac{1}{2}\left[1-\tanh\left(\frac{(C_n+C_h+C_d)-C^{(C)}}{\Delta c}\right)\right] : t\leq t^* \right\}. \label{chi^C}
   \end{equation}
\end{subequations}
Finally, the vasculature sensitivity threshold $\Delta c$ is estimated to be around $0.02$ $C^{(M)}$ cells.
 
\section{Modelling therapies}  
\label{sec:thetherapies}

\par

\subsection{Antithrombotic therapy}
\label{ATtherapy}
We will assume that low molecular weigh heparine (LMWH) decreases the expression of TF and increases the vascular functionality. Thus, it prevents the formation of coagulates. This therapy will be modelled by increasing the thresholds at which the vessels become unstable from $C^{(F)}=0.5C^{(M)}$ to $C^{(F)}=$ 0.6 - 0.8 $C^{(M)}$ and from $C^{(C)}=0.7 C^{(M)}$ to $C^{(C)}= 0.8-0.99 C^{(M)}$. Although the precise values are not known and probably depend on a number of additional features, they will allow us to simulate the proof-of-principle of the antithrombotic therapy.

\par

In addition to the prevention of VTE there are other several direct mechanisms of action of LMWH on tumor cells: direct cell killing \citep{Santos}, antiangiogenic effects \citep{Svensson} and many others \citep[see e.g.][Chap. 15]{Green}. Since those effects are difficult to quantify we will assume in this paper that the net effect of the antithrombotic therapy is to increase the vessel functionality thresholds $C^{(C)}$ and $C^{(F)}$.

\par

In this paper we will assume that drug administration leads to a blood concentration level that is sufficient to induce the required effect and as such 
we will not include the details of the pharmaco-kinetics of the drugs. The same applied to the treatment to be described in Sec. \ref{HIFtherapy}.

 \subsection{HIF-1$\alpha$ stabilization treatment}
 \label{HIFtherapy}
 
When oxygen is present, HIF-1$\alpha$ is continuously synthesized by the cell, but it is unstable and degraded with a half-life of about 2-3 minutes. However, as a consequence of the cycles of hypoxia due to the anomalous vasculature, free radicals are originated regulating and stabilizing the expression of HIF-1$\alpha$ even when vascular functionality is restored. In fact, U87 glioma cell line 
shows higher HIF-1$\alpha$ expression under cyclic than under chronic hypoxia \citep{Hsieh}. It is believed that the sequence of oxygen deprivation episodes may drive the accumulation of  HIF-1$\alpha$  in the cell nucleus. Therefore, cells need longer times to return to the normoxic phenotype state \citep{Semenza}. This HIF-1$\alpha$ stabilization significantly increases invasiveness and secretion of angiogenic factors and drives the intrinsic and extrinsic coagulation routes.
\par
Free radicals are usually detoxified by a complex system of proteins and antioxidant macromolecules 
which maintain the cell redox homeostasis. Antioxidative-based treatments, such as those with caulerpine or tempol, result in a HIF-1$\alpha$ inhibition and consequently in an antiproliferative effect in GBM murine xenograft models \citep{Hsieh}. Thus, antioxidative-based adjuvant therapies would provide faster ways for cells to revert to their normoxic state under oxic conditions. They decrease the HIF-1$\alpha$ accumulation and enable the recovery of the redox homeostasis. In our model, we will include the effect of the antioxidative therapy by decreasing the time of recovery of the normoxic phenotype $\tau_{hn}$ from 96 h to a smaller value in the range 8-48 h. 
\par
It is worth mentioning that based on the go or grow hypothesis, the diffusion coefficient of hypoxic cells is larger than for the normoxic cells, however, it is not known whether targeting  HIF-1$\alpha$ via antioxidative therapy would induce a significant change in them. Consequently, these parameters have a 10\% variability in each simulation and are constant in time along the treatments.

\subsection{Radiotherapy}
\label{XRT}
The standard radiotherapy protocol for high grade gliomas consists of a total of 60 Gy in fractions of 2 Gy given in 30 sessions from monday to friday leading to a treatment duration of 6 weeks. Treatment is  usually started several (2-4) weeks after surgery. 
\par
To describe the effect of the therapy in our model, cell death induced by radiation is included in the simulations instantaneously once per day (monday to friday) during 6 weeks.  We have chosen the simulations start 2 weeks after surgery, thus radiotherapy starts at week 2. 
\par
In the absence of oxygen, the unstable free radicals generated by radiation have a longer half life. Then, they can react with H$^+$ restoring its chemically original form without the need for biological and enzymatic intervention.
The overall result is that better oxygenated cells are more radiosensitive \citep{BCR}.
Thus we will assume the effect of radiotherapy to depend on the local oxygen concentration $O_2(x,t)$ at the time of the therapy. Denoting by $RT_{ox}$ the cell surviving fraction under oxic conditions and $RT_{hyp}$ the surviving fraction under hypoxic conditions, we will employ for the total surviving fraction under a local oxygen pressure $O_2(x,t)$, the equation
\begin{equation}
S (x,t) = RT_{ox} \frac{O_2(x,t)}{O_2^{v_i}} + RT_{hyp} \left(1-\frac{O_2(x,t)}{O_2^{v_i}} \right).\label{RT}
\end{equation}

\section{Parameter estimation and computational details}
\label{sec:theparameters}
\subsection{Parameter estimation}

\begin{table*} 
\caption{Typical values of the biological parameters taken for our model equations}
\label{fixparameters}       
\begin{tabular}{lllll}
\hline\noalign{\smallskip}
Parameter & Value and units &  Meaning  & Reference & Variation \\
\noalign{\smallskip}\hline\noalign{\smallskip}
$C^{(M)}$ &  $10^6$ cell/$cm^2$ & Maximum tumor cell density  & \cite{Swanson} & No \\
\noalign{\smallskip}\hline\noalign{\smallskip}
$O_2^{(S)}$ & 7 mmHg & Oxygen concentration level   & \cite{Vaupel2} & No  \\
                    &                & switch to hypoxia                   &                        &        \\
                    \noalign{\smallskip}\hline\noalign{\smallskip}
$O_2^{(T)}$ & 2.5 mmHg  & Michaelis Menten threshold  & \cite{Das} & No \\
 \noalign{\smallskip}\hline\noalign{\smallskip}
$O_2^{(D)}$ &  0.7 mmHg & Critical anoxia level  & \cite{Brown}  & No \\
\noalign{\smallskip}\hline\noalign{\smallskip}
$O_2^{v}$ &  80 mmHg & Typical oxygen pressure & \cite{Kimura} & No \\
                  &                   & within vessels                 &                      &   \\
                  \noalign{\smallskip}\hline\noalign{\smallskip}
$\alpha_n$ & $7.5 \times 10^{-4}$  & Typical normoxic cell  &  \cite{Das} & No \\ 
                   &       mmHg c/s                                            & oxygen consumption  &                   &      \\
                   \noalign{\smallskip}\hline\noalign{\smallskip}
$\alpha_h$ &  $\alpha_n/5$ mmHg c/s & Typical hypoxic cell  & \cite{Griguer08}  & No \\
                   &                                         & oxygen consumption  &   & \\
                   \noalign{\smallskip}\hline\noalign{\smallskip}
$D_{O_2}$ &  $10^{-5}$ cm$^2$/s & Oxygen diffusion coefficient & \cite{Das} & No \\
\noalign{\smallskip}\hline\noalign{\smallskip}
$\tau_{nh}$&  $0.15$ h & Normoxic to hypoxic   & \cite{Jewell}  & No \\
                  &                  & phenotype switch time  & & \\
                  \noalign{\smallskip}\hline\noalign{\smallskip}
$\tau_{hd}$ & 7 days & Anoxic death time & \cite{Alicia} & No \\
\noalign{\smallskip}\hline\noalign{\smallskip}
$J_{O_2}$ & $10^{-6}$ s$^{-1}$ & Oxygen flow coefficient & Estimated & No\\
\noalign{\smallskip}\hline\noalign{\smallskip}
$\Delta c$ & $0.02$ $C^{(M)}$ cells & Sensitivity vascular  threshold & Estimated & No  \\ \noalign{\smallskip}\hline\noalign{\smallskip}
$D_n$ &  $5\times10^{-10}$   cm$^2$/s & Normoxic and hypoxic &  & 10\% \\
$D_h$ &  $5 \times 10^{-9}$  cm$^2$/s  &  diffusion coefficients & \cite{Swanson}   & 10\% \\ \noalign{\smallskip}\hline\noalign{\smallskip}
$\tau_n$ & 14 days & Normoxic and hypoxic  & Estimated by Fisher- &  10\%\\
$\tau_h$ & 24 days & doubling times  & Kolmogorov approx. & 10\%\\ \noalign{\smallskip}\hline\noalign{\smallskip}
$\tau_{hn}$ & 96 hours & Hypoxic to normoxic   & \cite{Semenza}& 10\%\\
& & phenotype switch time & \cite{Hsieh} & \\ \noalign{\smallskip}\hline\noalign{\smallskip}
$C^{(F)}$ &$0.5$ $C^{(M)}$& Critical cell number  & \cite{Ruf} & 10\%\\
                                 &       & for fluctuating hypoxia &       & \\ \noalign{\smallskip}\hline\noalign{\smallskip}
$C^{(C)}$&$0.7$ $C^{(M)}$& Critical cell number &  \cite{Ruf}& 5\%\\
                                   &     & for chronic hypoxia &       & \\ \noalign{\smallskip}\hline\noalign{\smallskip} 
$\tau_{hn}$ & 8-48 hours & Hypoxic to normoxic  & \cite{Semenza} & 5\% \\
		&                    &	phenotype switch time						&   &\\
                   &                   &   under AO therapy & \cite{Hsieh} & \\ \noalign{\smallskip}\hline\noalign{\smallskip} 
$C^{(F)}$ &$0.6$-$0.8$ $C^{(M)}$& Cell number thresholds & \cite{Ruf} & 5\%\\
$C^{(C)}$&$0.8$-$0.99$ $C^{(M)}$                             &   for fluctuating and chronic &       & 1\% \\
                                &        &  hypoxia under AT &       & \\
                                 \noalign{\smallskip}\hline\noalign{\smallskip}
$RT_{ox}$ & $0.8$ & RT surviving fractions  & \cite{BCR}& 5\%\\
$RT_{hyp}$& $0.95$& in normoxia and hypoxia &  & 5\%\\
\noalign{\smallskip}\hline
\end{tabular}
\end{table*}

We resort to available experimental values from human glioma models to obtain order-of-magnitude estimates of the intervening parameters in our equations. 
Typical values of the used biological parameters are shown in Table. \ref{fixparameters} (see also further parameters details in \citet{Alicia}).
\par
First, the maximum cell density $C^{(M)}$ has been estimated in previous works \citep[see e.g.][]{Swanson} to be about $10^6$ cell/cm$^2$. Oxygen pressure threshold for hypoxic metabolism $O_2^{(S)}$ is cell line dependent but experimental evidence supports for glioma the choice of 7 mm Hg \citep{Vaupel2}. The Michaelis-Menten constant, $O_2^{(T)}$ has to be smaller than this parameter yet larger than anoxia threshold, $O_2^{(D)}$, about 0.7 mm Hg \citep{Brown}. We have chosen it to be 2.5 mm Hg \citep{Das}. 
The oxygen diffusion coefficient $D_{O_2}$ is classically known to be around $10^{-5}$ cm$^2$/s \citep{Das} while the cell diffusion coefficients are not so readily accessible {\em in vivo}. 
\par
 The hypoxic diffusion coefficient has been considered to be around the mean data reported in \citet{Swanson} and ten times larger than the normoxic one. 
In addition, for high-grade glioma, the largest radial velocity $v$ ($\approx$3 cm/y) can be related to the hypoxic cell propagation whereas the smallest radial velocity ($\approx$1 cm/y) can be related to the normoxic one. For instance, considering the Fisher-Kolmogorov approximation to the radial cell propagation ($v\approx 2\sqrt{D/\tau}$),
 the normoxic and hypoxic doubling times $\tau$ (14 and 24 days respectively) can be extrapolated, which are also similar to those used in \cite{Swanson}. The normoxic oxygen uptake was obtained from a healthy tissue of $0.2C^{(M)}$ as cell density with oxygen consumption of 15 mmHg/s \citep{Das}. Hypoxic oxygen uptake is several times smaller than the normoxic one as it was observed in U251 glioma cells \citep{Griguer08}.

\subsection{Computational details}

Henceforth, we consider two possible scenarios: one corresponding to a biopsied but not totally/partially resected tumor, and a second one where the tumor was either partially or totally resected. In both scenarios we will assume that, prior to the application of radiotherapy, or antithrombotic (AT) and antioxidative (AO) therapies, there is a mixture of normoxic cells plus smaller fractions of hypoxic cells and a necrotic core. 

In the case of resected tumors initially there will be a small infiltrative tumor remnant located around the surgical border.
In that case simulations start two weeks after surgery and all blood vessels are assumed to work correctly, thus $\chi^C=\chi^F=1$. Since the initial cell density is low, there is a good oxygen supply at $t=0$ and its redistribution is very fast, we can take $O_2(x,0)$ to be rougly uniform in space. 
\par

Non-resected tumors will be considered of size $\approx$ 1.6 cm of diameter at diagnosis, with a necrotic core at its center occupying  $\approx$ 0.8 cm of diameter. Around the necrotic tissue a high tumor cell density of about 0.6 (in units of $C^{(M)}$) exists and is formed by normoxic and hypoxic cells. This hypercellular region is related to the high-contrast ring which is frequently observed in T1 MRIs (with Gadollinium) of GBMs. In our simulations, the width of this ring is about 4 mm. A very low density of hypoxic cells infiltrating into the healthy tissue (brain parenchima) is also present. Only blood vessels located within the necrotic compartment are considered to be non functional.   We will also study smaller tumors for partially resected tumors, of diameter $\approx$ 1 cm after surgery, where the necrotic core is in the center occupying  $\approx$ 0.2 cm of diameter and radial tumor infiltration of size $\approx$ 0.4 cm formed by normoxic and hypoxic cells. In that situation we will assume that there are no functional vessels in the center of the tumor.

 \par
 
To solve Eqs. \eqref{themodel} and \eqref{oxygen} numerically we have used a standard finite difference method of second order in time and space with zero boundary conditions along the sides of the computational domain. We have employed large integration domains and cross-checked our results for different domain sizes to avoid spurious edge effects. All the results provided in the following sections are the outcome of sets of ten simulations. Each simulation allows a random static variability (in the range of 1-10$\%$) in the parameters according to the last column in table \ref{fixparameters}. All variations are constant along time except $z$ in Eq. \ref{chi^F}, which is related to the stochastic temporal variability of oxigen perfusion from the vessels. The error bars shown in the figures correspond to the standard deviations of the results after the full set of simulations corresponding to small variations of the parameters is computed.

\section{Results}
\label{sec:theresults}

\subsection{Tumor evolution}

\begin{figure*}
\begin{center}
\epsfig{file=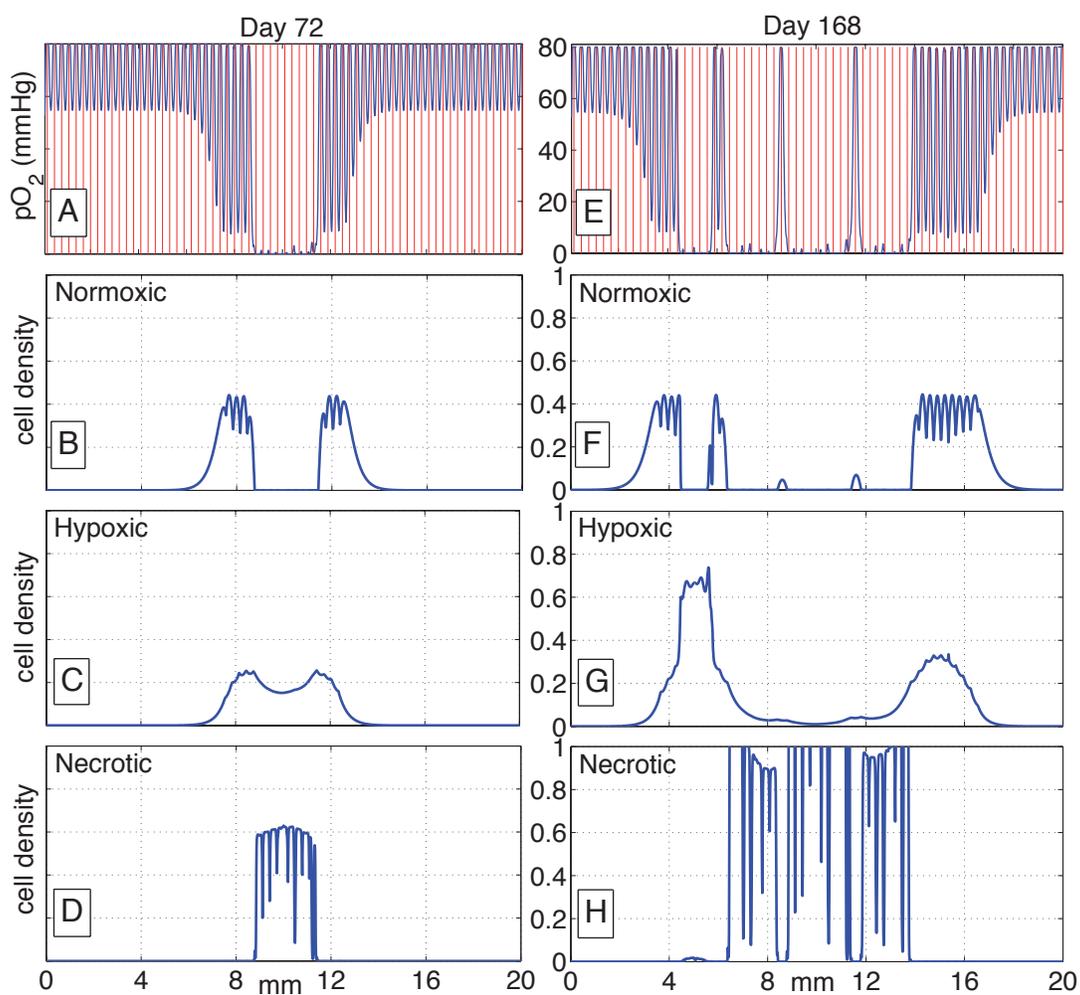,scale=0.5} 
\end{center}
\caption{{\bf Typical evolution of the cell densities and oxygen on a tissue of size 2 cm.} Intermediate distribution (day 72) on the left column and final distribution (day 168) on the right. A and E display tissue oxygenation, B and F the normoxic cell density, C and G the hypoxic cell density and finally, D and H show the necrosis. Red vertical lines at A and E simulate the blood vessels position along the tissue.  \label{fig:CI}}
\end{figure*}

We have performed extensive simulations of the model equations (\ref{themodel}-\ref{chis}) with different initial data and for broad parameter ranges of biological significance. A typical result of the tumor density evolution is shown in Fig. \ref{fig:CI} for two times (days 72 and 168). In spite of the extensive cytoreduction assumed for the surgery, tumor cells proliferate giving rise to a relapse. The first stages (not shown in the figure) correspond to the proliferation of the normoxic component until a sufficiently high density occurs that leads to vessel instability as described in Sec. \ref{vascu}. After 72 days, a number of vessels in the center of the tumor have already collapsed developing anoxic areas as shown in the oxygen distribution (Fig. \ref{fig:CI}A). As a result, a significant fraction of hypoxic cells start to coexist and compete with the normoxic ones (Fig. \ref{fig:CI}B). Finally, a necrotic core of about 3 mm in size occupies the center of the tumor (Fig. \ref{fig:CI}D). Due to the effect of the tumor on the  vessel functionality and the differences in the cell uptake coefficients, there is a significant dispersion of oxygenation levels in regions containing high cell densities. 

\par
As the tumor progresses, the area occupied by damaged vessels and necrosis increases with time,  and oxygenation decreases due to the destruction of the functional vasculature (Fig. \ref{fig:CI}E). After 168 days the necrotic core already has a size of 8 mm (Fig. \ref{fig:CI}H).  Necrotic areas are typically surrounded by hypercellular regions generated by the migration of cells from low oxygen areas, resembling the pseudopalisades observed in high-grade gliomas \citep{pseudopalisading1,pseudopalisading2}. 

\begin{figure*}
\begin{center}
\epsfig{file=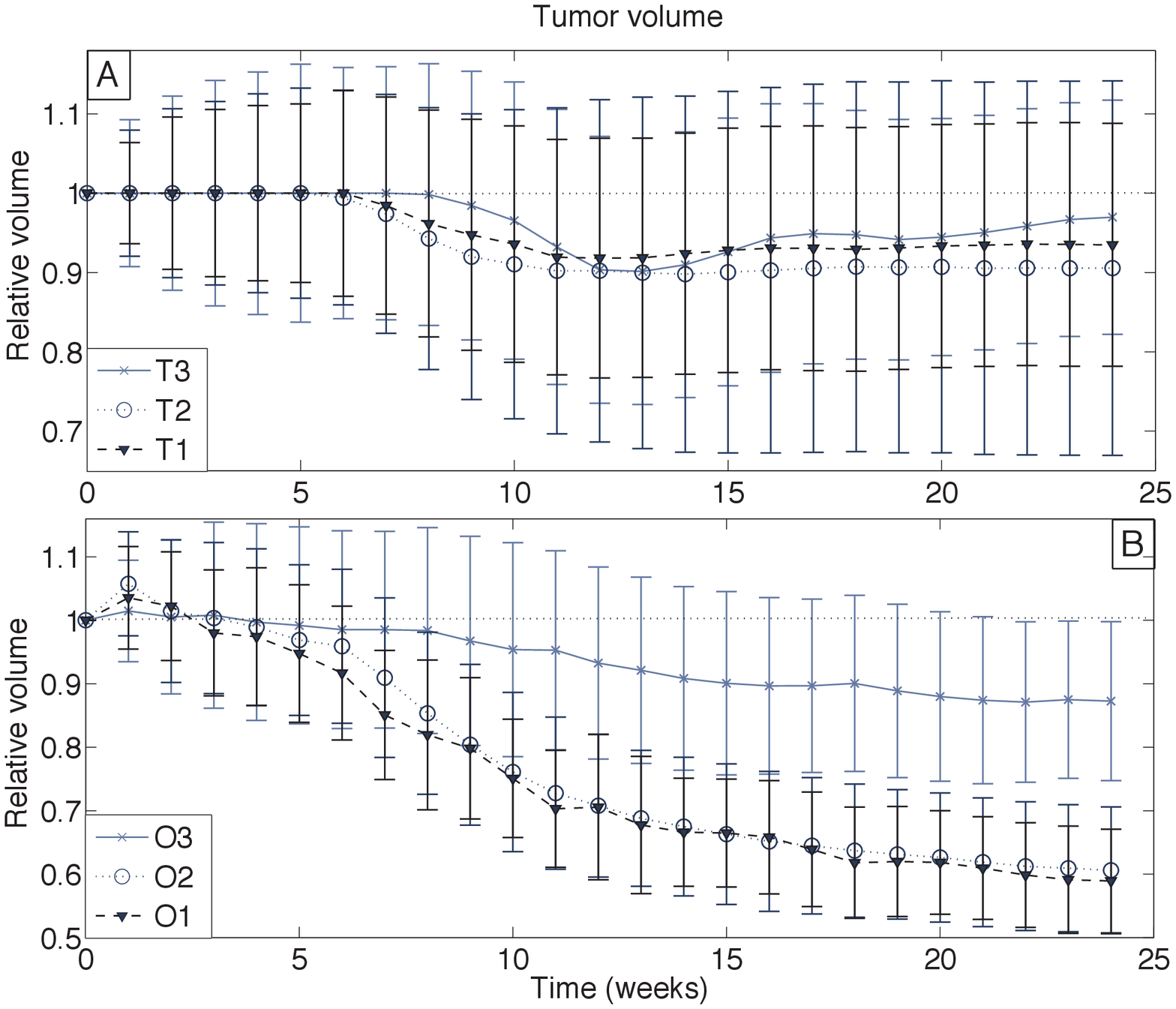,scale=0.38} 
\end{center}
\caption{{\bf Volume of tumors treated with antithrombotics or antioxidants related to non treated tumors.} A: Relative tumor volume for different antithrombotic regimens (T3: $C^{(C)}=0.8 C^{(M)}$, $C^{(F)}=0.6 C^{(M)} $ , T2: $C^{(C)}=0.9 C^{(M)}$, $C^{(F)}=0.7 C^{(M)}$ , T1: $C^{(C)}=0.99 C^{(M)}$, $C^{(F)}=0.8 C^{(M)})$. B: Relative tumor volume for different antioxidative regimens (O3: $\tau_{hn}= 48$ hours O2: $\tau_{hn}= 24$ hours  O1: $\tau_{hn}= 8$ hours). Doted lines in A and B are the respective relative volumes for non-treated tumors.  Error bars display the standard deviation for the 10 simulations developed for each treatment varying the parameters as indicated in Table \ref{fixparameters}. \label{fig:relative_volumen}}
\end{figure*}

\subsection{Targeting vaso-occlusions leads to a small reduction in tumor volume}
\label{voccl}

We have first studied the effect of  antithrombotic (AT) therapy improving vessel functionality as described in Sec. \ref{ATtherapy}. This therapy reduces invasion speed by keeping cells in their oxic and less motile phenotype but the effect on tumor progression is small. An example is shown in Fig. \ref{fig:relative_volumen}A. Each point represents the average of 10 simulations with mean values and static random variations (Table  \ref{fixparameters}). The error bars indicate the standard deviation of the results of those sets of simulations. 
\par
We have used different antithrombotic doses T1, T2, and T3 
to delay vessel degradation by the tumor. We suppose that there is a dosis-response effect. The threshold parameters for chronic and fluctuating hypoxia  for T3 are $C^{(C)}=0.8 C^{(M)}$ and  $C^{(F)}=0.6 C^{(M)} $, for T2: $C^{(C)}=0.9 C^{(M)}$, $C^{(F)}=0.7 C^{(M)}$ and for T1: $C^{(C)}=0.99 C^{(M)}$, $C^{(F)}=0.8 C^{(M)}$. It is clear from Fig. \ref{fig:relative_volumen} that there are no significant differences with the non-treated tumor volumes until the sixth week; this is the time were the tumor density becomes large enough to impair the vessels. This therapy is able to delay the damage of the vessels but once the new (higher) thresholds are reached no further gains are observed. We have verified this conclusion with a large range of parameters obtaining very similar results. Thus, only minor gains are to be expected from AT therapy unless it is combined with other agents. 
\par

\subsection{Antioxidative therapy has the potential to improve survival substantially}
\label{THIF}
The heterodimer HIF-1 has complex effects on tumor cell biology. On the one hand, the suppression of the HIF-1$\alpha$ subunit activity severely compromises the ability of tumor cells to undergo anaerobic glycolysis. This reduces the proliferation rate of hypoxic cells and promotes the apoptotic program in cells that are deprived of both oxygen and glucose \citep{BCR}. This is why many therapies have tried to target HIF-1$\alpha$, although their effectivity in brain tumors has not been proven yet. In our case we propose antioxidative therapies as a way to normalize the cell response under oxic conditions.  
\par

We have used different antioxidant doses O3, O2 and O1, corresponding to a reduction in the normalization times to  $\tau_{hn}= 48, 24,$ and 8 hours respectively. Our simulations encompass a time window of six months for the tumor volume evolution with and without AO therapies. The results are depicted in Fig. \ref{fig:relative_volumen}B.  Since the AO therapy promotes the proliferative phenotype, vaso occlusion takes place earlier than in the control case. This leads to the formation of cells actively migrating away from collapsed vessels and the AO therapy does not decrease the tumor volume during the first 4 weeks. However, maintenance of the therapy for longer times produces a significant benefit coming from the essential lessening in the invasion speed. Even assuming a minor effect of the AO therapy (as would be the case with O3) there is a significant tumor volume reduction (12\%) after six months.  Higher doses/effects of antioxidants (O1 or O2), associated to smaller recovery times, result in substantial tumor volume reductions (up to 40\%) (Fig. \ref{fig:relative_volumen} B). We have used a large range of parameters obtaining similar results that would suggest that a prolonged AO therapy may be beneficial for a large number of patients.

\subsection{Antithrombotics and antioxidants may have a synergistic effect leading to a substantial tumor volume reduction}
Having AT and AO together in our mathematical model, has the biological meaning that we are simultaneously targeting the coagulation cascade to delay the invasiveness process and the high levels of free radicals that also have an effect on invasiveness.
\par
\begin{figure}
\begin{center}
\epsfig{file=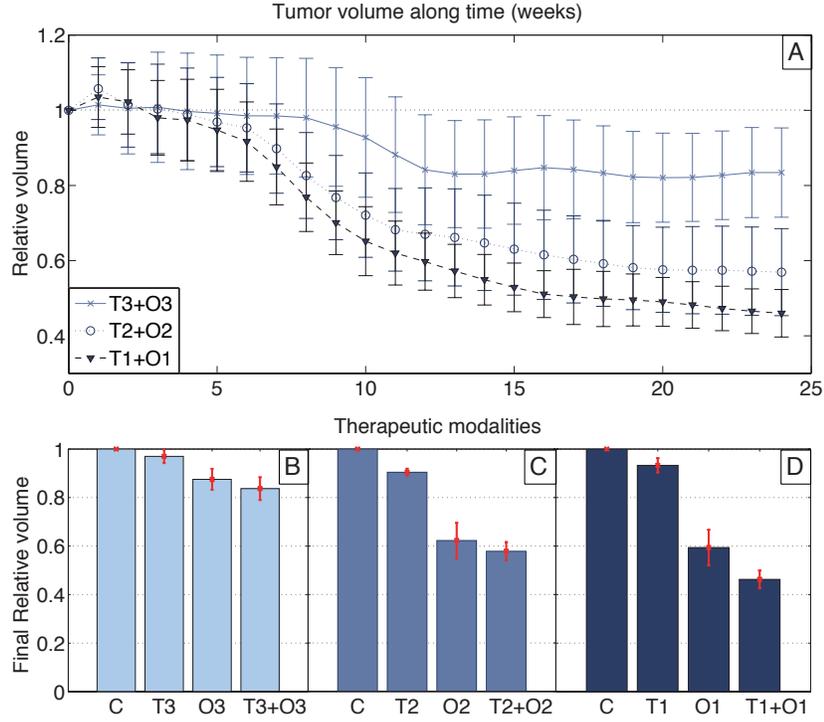,width=11cm} 
\end{center}
\caption{{\bf Evolution in time of the relative tumor volume under combination therapies with antithrombotics and antioxidants.} A: Tumor volume as a function of time for the three combined treatments relative to the control. The dotted constant line represents the volumes for non treated tumors. Antithrombotic treatments are T3: $C^{(C)}=0.8 C^{(M)}$, $C^{(F)}=0.6 C^{(M)} $, T2: $C^{(C)}=0.9 C^{(M)}$, $C^{(F)}=0.7 C^{(M)}$ and T1: $C^{(C)}=0.99 C^{(M)}$, $C^{(F)} =0.8 C^{(M)}$. Antioxidative treatments are  O3: $\tau_{hn}= 48$ hours, O2: $\tau_{hn}= 24$ hours and  O1: $\tau_{hn}= 8$ hours.  B, C, D: Final relative tumor volumes for the different therapeutic modalities considering C as the control (non treated tumor group).  
 Error bars display the standard deviation for the 10 simulations developed for each treatment varying the parameters as indicated in Table \ref{fixparameters}.
 \label{fig:volumencombinado}}
\end{figure}

Fig. \ref{fig:volumencombinado} shows the typical outcome of our simulations. We have combined the therapies T1, T2, T3, O1, O2 and O3, as described in  Secs. \ref{voccl} and \ref{THIF}, to get {\em in silico} several treatment modalities: T3+O3, T2+O2 and T1+O1, besides the control C, which is the non treated tumor group.  The error bars in Fig. \ref{fig:volumencombinado} display the standard deviation for ten simulations varying the parameters in Table \ref{fixparameters}, as described previously.
\par
Firstly, Fig. \ref{fig:volumencombinado}A exhibits the tumor volume evolution related to the control for the combined treatments. All treatments show a visible volume reduction from the sixth week. After 24 weeks, the relative tumor volume reduction, as compared to the control, is 0.83, 0.56 and 0.45 for T3+O3, T2+O2 and T1+O1, respectively. 

Secondly, Fig. \ref{fig:volumencombinado} subplots B, C and D  show  the tumor volumes after six months for each different therapeutic modality. All of them result in a  reduction in the final  tumor volume after six months of treatment, but in all cases the combination of AT and AO gives the best outcome. In some instances, such as with the most effective therapies T1+O1, a synergistic effect  is observed  when combining both therapies (see Fig. \ref{fig:volumencombinado}D).

\subsection{Antithrombotics and antioxidants sensitize the tumor for cytotoxic therapies}

An interesting result of our simulations is that the combination of antithrombotics and antioxidants results in a reduction of hypoxic areas and the normalization of cell phenotypes that results in a higher sensitivity to cytotoxic therapies. In what follows we will discuss the action of radiotherapy (RT) together with the AT and AO therapies. As discussed in Sec. \ref{XRT}, well oxygenated cells suffer more damage when exposed to radiation. In addition, more aggressive phenotypes, such as those induced by hypoxia, have higher levels of repair enzymes that makes them less sensitive to cytotoxic therapies.

\par

\begin{figure*}
\begin{center}
\epsfig{file=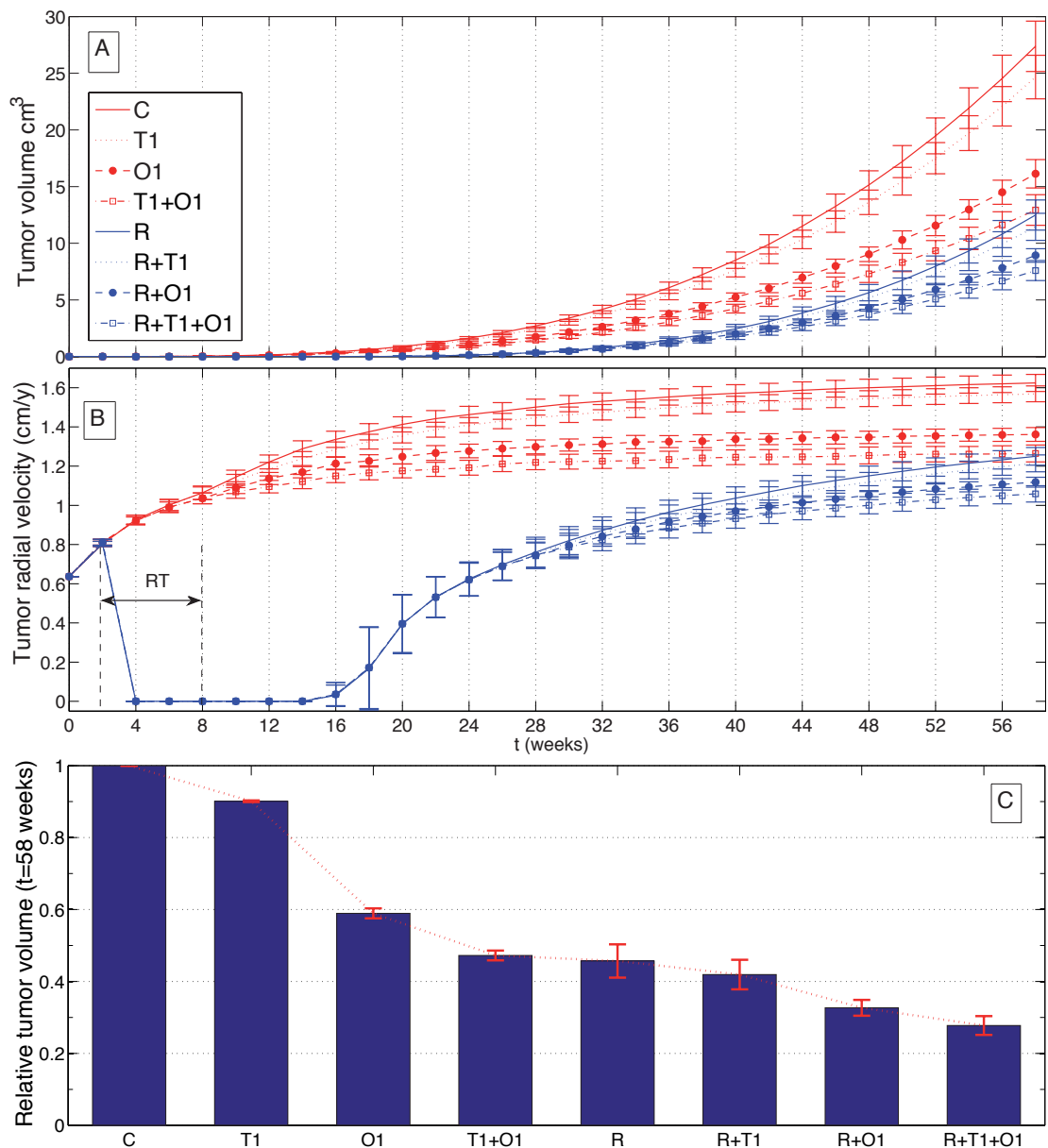,scale=0.46} 
\end{center}
\caption{{\bf Tumor volume and tumor radial velocity for different therapeutic modalities.} Subplots A and B display the tumor volume (cm$^3$) and the tumor radial velocity (cm/y) during 58 weeks, respectively. Each line provides the results for a therapeutic modality:  C is the control (non-treated),  T1 (antithrombotics): $C^{(C)}=0.99 C^{(M)}$, $C^{(F)}=0.8 C^{(M)}$,  O1 (antioxidants): $\tau_{hn}= 8$ hours, T1+O1 (antithrombotics+antioxidants), R (standard radiotherapy  with 30 sessions, 2 Gy/session, monday-friday), R+T1 (radiotherapy+antithrombotics), R+O1 (radiotherapy+antioxidants),  R+T1+O1 (radiotherapy+antithrombotics+antioxidants). Red curves correspond to non-radiated tumors and blue curves to radiated ones. Subplot C shows the relative final volume for all treatments after 58 weeks. Error bars display the standard deviation for the 3 simulations developed for each treatment varying the parameters as indicated in Table \ref{fixparameters}. \label{fig:velocity}}
\end{figure*}

We have run a set of simulations combining RT with AT and AO for tumors of different initial size at diagnosis. We display examples of their evolution, comparing different combination therapies and initial conditions: totally resected tumors after surgery (Fig. \ref{fig:velocity}),  partially resected tumors (Fig. \ref{fig:partialresected}) and biopsied tumors without an extensive tumor volume reduction (Fig. \ref{fig:biopsy}).

\par

A first example of our results is plotted in Fig. \ref{fig:velocity} where we show the {\em in silico} predictions for the tumor volume evolution (Fig. \ref{fig:velocity}A), radial tumor velocity (Fig. \ref{fig:velocity}B), for different therapeutic modalities during a period of 58 weeks, together with the final volumes comparing all therapies Fig. \ref{fig:velocity} C. Following the standard practice for GBM we will assume that RT starts two weeks after surgery and runs  for 6 weeks from monday to friday in fractions of 2 Gy. It is clear from Fig. \ref{fig:velocity}A and B that there are no significant volume or growth speed differences between the irradiated tumors until week 12. The effect of RT results in a substantial reduction in tumor volume and tumor growth speed that is still apparent several months after the therapy has finished. It is also obvious that combination therapies of RT with the different agents are always more effective than radiotherapy alone. It is remarkable that the effect of AT+AO is equivalent, after one year, to the effect of RT, though with a much lower toxicity (and long-term effects). At the end of the studied period of 58 weeks, there is a $12\%$ reduction for AT alone and a remarkable $42\%$ decrease with AO alone. The mean final tumor volume for the combination of T1+O1 is similar to the one obtained for tumors receiving only RT; both displaying a reduction of more than 50\% when compared to the control group.
RT+AT gives rise to about 55\% reduction, RT+AO to around 65\% and the combination RT+AT+AO exhibits over 70\% of tumor volume at week 58th compared to the non-treated tumors. 

\par

Our results suggest that a substantial increase of several months in the median overall survival for GBM patients may be obtained by combining RT+AT+AO.  
\par

\subsection{Antithrombotics and antioxidants might be more effective for patients having only a biopsy or subtotal resections}
\begin{figure*}
\begin{center}
\epsfig{file=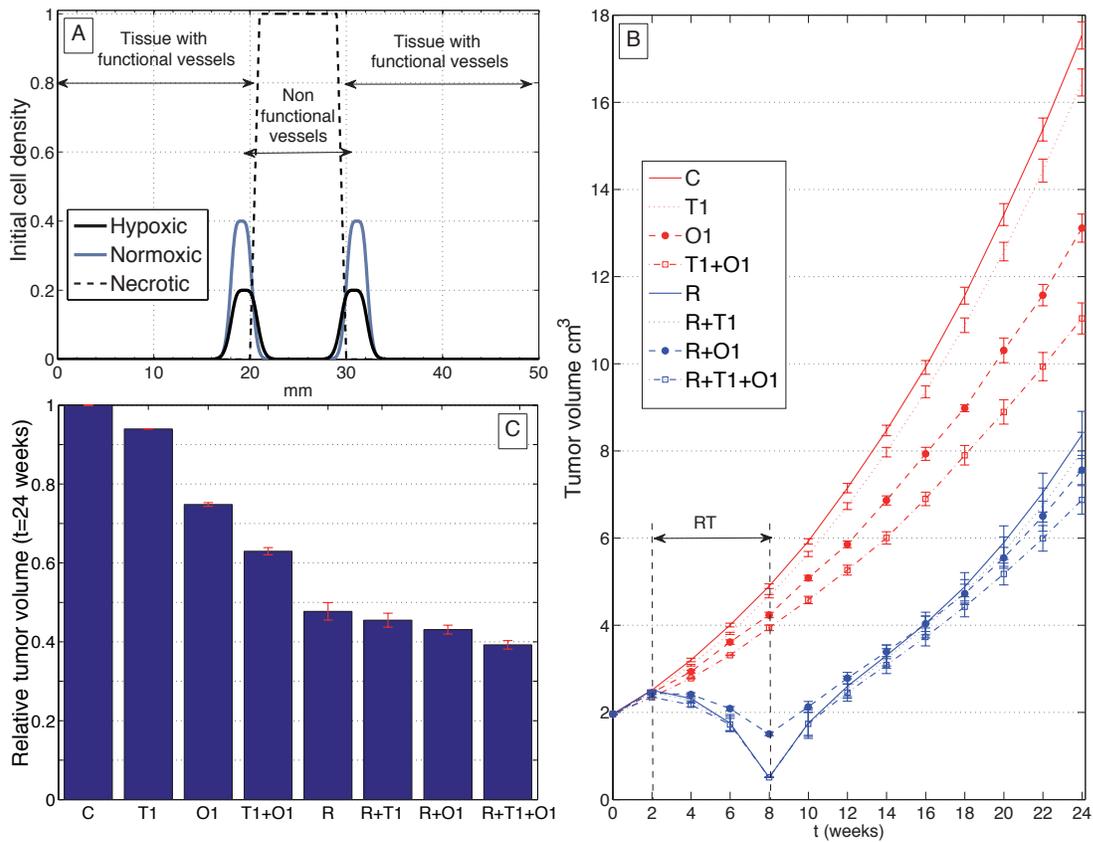,scale=0.41} 
\end{center}
\caption{{\bf Tumor volume evolution for different therapeutic modalities without prior surgery.} Subplot A shows the initial cell density distribution at diagnosis for hypoxic, normoxic and necrotic cells. Initial tumor diameter $\approx$ 1.6 cm, necrotic core $\approx$ 0.8 cm and radial tumor infiltration $\approx$ 0.5 cm. Subplot B displays the tumor volume evolution (cm$^3$) during 24 weeks. Each line provides the results for a therapeutic modality:  C is the control (non-treated),  T1 (antithrombotics): $C^{(C)}=0.99 C^{(M)}$, $C^{(F)}=0.8 C^{(M)}$,  O1 (antioxidants): $\tau_{hn}= 8$ hours, T1+O1 (antithrombotics+antioxidants), R (standard radiotherapy with 30 sessions, 2 Gy/session, monday-friday), R+T1 (radiotherapy+antithrombotics), R+O1 (radiotherapy+antioxidants),  R+T1+O1 (radiotherapy+antithrombotics+antioxidants). Red curves correspond to non-radiated tumors and blue curves to radiated ones. Subplot C shows the relative final volume for all treatments after 24 weeks. Error bars display the standard deviation for the 3 simulations developed for each treatment varying the parameters as indicated in Table \ref{fixparameters}. \label{fig:biopsy}}
\end{figure*}

Fig. \ref{fig:biopsy} displays the tumor volume evolution for the different therapeutic modalities during 24 weeks after diagnosis. In this case, the tumor was not resected and its diameter at diagnosis was higher than 1.6 cm. Only vessels located between 20 and 30 mm were considered non functional since almost this space was occupied by dead cells as illustrated in the initial cell density subplot Fig. \ref{fig:biopsy}A.  Tumor volume evolution is shown  in  Fig. \ref{fig:biopsy}B and evidences how all therapies delay tumor progression almost from the first day, a fact that did not occur for totally resected tumors. 
\par

\par
At the end of the 24 weeks, we find a benefit in the reduction of tumor volume similar to the one obtained from totally resected tumors after 58 weeks of treatment modalities  (Fig. \ref{fig:biopsy}C). Thus, the group of patients with biopsy or subtotal resection might benefit even more from  a combination of AT and AO therapies. In addition, when both AT and AO are used together with radiotherapy a significant advantage is obtained when compared with radiotherapy alone. This gain is more evident for patients having only a biopsy or subtotal resections, than in tumors with small remnants that are expected to be initially well oxygenated.

\par

\subsection{Combination of antithrombotics and antioxidants might be more effective than radiotherapy alone}

\begin{figure*}
\begin{center}
\epsfig{file=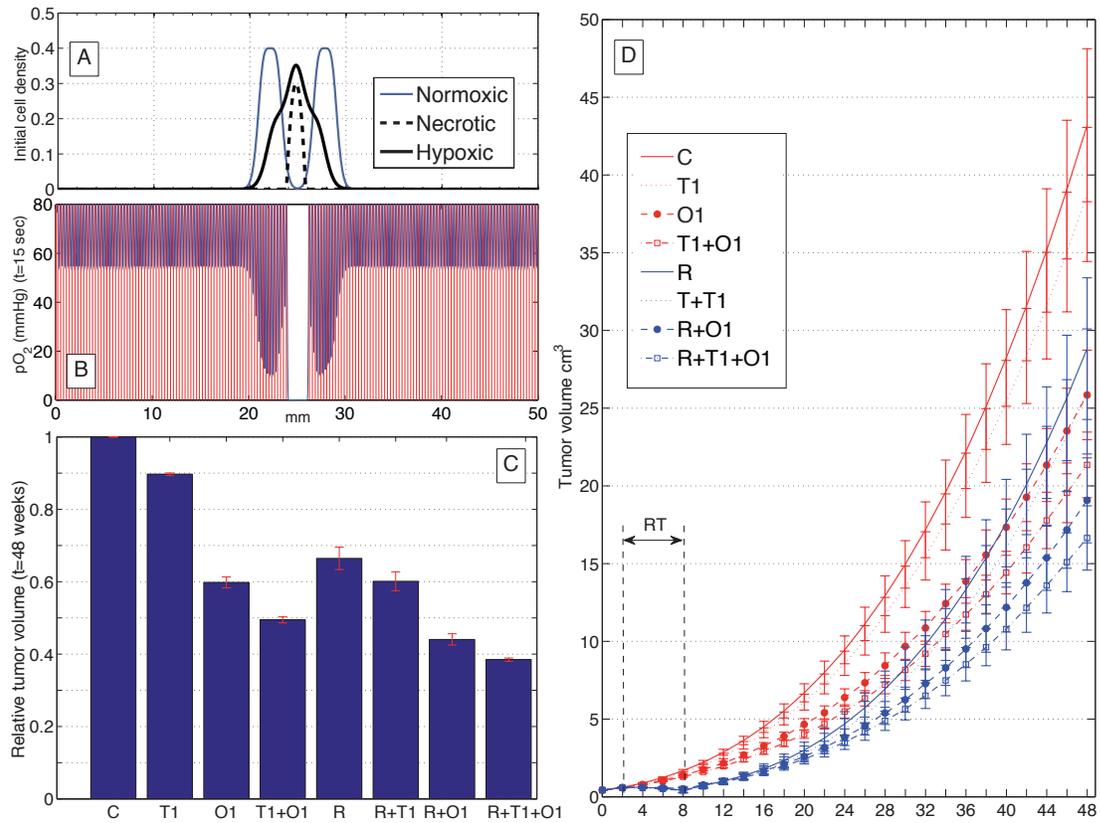,scale=0.41} 
\end{center}
\caption{{\bf Tumor volume evolution for different therapeutic modalities after subtotal resection.} Subplot A shows the initial cell density distribution for hypoxic, normoxic and necrotic cells. Initial tumor diameter $\approx$ 1 cm, necrotic core $\approx$ 0.2 cm and radial tumor infiltration $\approx$ 0.4 cm. B displays the initial distribution of the partial pressure of oxygen. Vertical red lines represent the functional vessels and blue lines the oxygenation along the space. There are no functional vessels at the center of the tumor. Subplot C shows the relative final volume for all treatments after 48 weeks. Subplot D displays the tumor volume evolution (cm$^3$) during 48 weeks.
 Each line provides the results for a therapeutic modality :  C is the control (non-treated),  T1 (antithrombotics): $C^{(C)}=0.99 C^{(M)}$, $C^{(F)}=0.8 C^{(M)}$,  O1 (antioxidants): $\tau_{hn}= 8$ hours, T1+O1 (antithrombotics+antioxidants), R (standard radiotherapy with 30 sessions, 2 Gy/session, monday-friday), R+T1 (radiotherapy+antithrombotics), R+O1 (radiotherapy+antioxidants),  R+T1+O1 (radiotherapy+antithrombotics+antioxidants). Red curves correspond to non-radiated tumors and blue curves to radiated ones. Error bars display the standard deviation for the 3 simulations computed for each treatment varying the parameters as indicated in Table \ref{fixparameters}. \label{fig:partialresected}}
\end{figure*}

The last example shown in Fig. \ref{fig:partialresected} displays the tumor volume evolution for the different therapeutic modalities during the first 48 weeks after subtotal resection. In this case, the tumor diameter following surgery was around 1 cm. Initially, the vessels in the center of the tumor were taken to be non functional leading to the formation of a small necrotic core. Hypoxic and normoxic cells, surrounding the necrotic areas, infiltrate into the healthy tissue Fig. \ref{fig:partialresected}A. The initial distribution of the partial pressure of oxygen is displayed in Fig. \ref{fig:partialresected}B. 
\par
After 48 weeks all treated tumors significantly reduced their volume compared to the non treated ones. It is remarkable that radiotherapy implies a 35\% reduction versus the 55\% reduction observed in Fig. \ref{fig:velocity}C for totally resected tumors. The cause of this fact can be the better initial oxygenation and the absence of necrosis and hypoxic cells in the second case. However, antioxidants and antithrombotics are still as effective as in Fig. \ref{fig:velocity}. In addition, after 48 weeks, the benefit obtained from the combination of AT and AO with radiotherapy is more apparent for these cases (see Fig. \ref{fig:partialresected}C).
\par
Although the best result is obtained with a combination of the three proposed therapies, it is important to emphasize the fact that AO alone or AO with AT (without radiation) produced a  reduction in tumor volume larger than the one observed under radiotherapy alone (see Fig. \ref{fig:partialresected}D).
\par

\section{Discusion and conclusions}
\label{sec:theconclusions}

As a consequence of cyclic hypoxia during GBM progression, free radicals are mainly originated from the mitochondrial respiratory chain which regulates the transcription hypoxia-inducible factor HIF-1$\alpha$ expression. Indeed, hypoxia (particularly via HIF-1$\alpha$) is responsible  for metabolic and phenotypic changes which induce invasiveness of GBM cells and initiate coagulation mechanisms that promote tumor growth acceleration.  
\par
 The antioxidative treatment  may result in the inhibition of HIF-1$\alpha$ leading to a reduction of the tumor invasiveness, delaying progression of GBM in humans. In addition, the increase of free radical could be stabilized by means of an antioxidative therapy, controlling or delaying the vascular degeneration due to the malformations of endothelial cells induced by the HIF-1$\alpha$ pathway activation.  In this way, targeting HIF-1$\alpha$ can be seen as an attractive approach to complement radiotherapy and chemotherapy, which kill well-oxygenated cells (\cite{BCR, Chen}) and also to inactivate extrinsic HIF-1$\alpha$ pathways, forcing the tumor to display a lower-grade behavior by reducing invasion and angiogenesis. 
 \par
 Other potential interesting agents for GBM treatment are antithrombotics, such as low molecular weight heparine (LMWH), a low molecular weight heparin that promotes the release of tissue factor pathway inhibitors for preventing venous thromboembolism and with very low toxicities \citep{Planes}. Thus LMWH might be incorporated as a part of GBM treatment to avoid vaso-occlusion phenomena and to delay  the neovascularization process associated with this pathology. The results from our model seem to imply that an additional indirect antitumoral effect from thromboprophylaxis might be expected related to the delay of tumor invasion, different from the direct antitumoral effect or the increase in survival and quality of life to be expected from the prevention of the formation of big coagulates. Furthermore, from our simulations shown in Fig. \ref{fig:relative_volumen}A, it follows that an optimum antithrombotic treatment regimen exists (in our case, T2: $C^{(C)}=0.9 C^{(M)}$, $C^{(F)}=0.7 C^{(M)}$ in comparison with T3 or T1) as evidenced by a U-effect on the relative tumor volume. 
 
 \par
In addition, antithrombotic therapy could avoid early tumor-induced vaso occlusions delaying the so-called malignant transformation of low-grade gliomas, were they transition to a high-grade glioma. Consequently, antithrombotic therapy administered to WHO grade II astrocytic and oligodendroglial patients may help to preclude the malignant transformation into secondary GBM associated with the onset of hypoxia once local vascular damage is produced. 

\par
Finally, Targeting HIF-1$\alpha$ and vasculature normalization simultaneously in GBM may have a synergistic effect decreasing tumor invasion and increasing patient overall survival time by changing the tumor microenvironmental behavior from high grade to low grade glioma. In addition, tumor cells have high levels of antioxidative enzymes which contribute to eliminate more efficiently the free radicals produced by alkylating agents such as carmustine (BCNU) or by the radiation. Besides the radiotherapy sensitization obtained with a better tumor oxygenation, the treatment would also reduce the need to activate antioxidative enzymes, rendering the tumor more susceptible to the effect of these agents. Fig. \ref{fig:therapies} summarizes the possible synergetic effect predicted by our \emph{in silico} model combining antioxidants, antithrombotics and radiotheraphy in the GBM treatment.
\par

It is worth to mention the use of anti-angiogenics in the context of the go or grow hypothesis. On the one hand, anti-angiogenic treatments impair tissue oxygenation and promote motile phenotypes to dominate the tumor population and invade faster the adjacent tissue. On the other hand, a pro-vascular therapy may oxygenate the tumor with the risk of producing leaky vessels through which tumor cells can intravasate and form distant foci in the brain parenchyma. This has lead to a long discussion on the potential existence of a window where anti-angiogenic therapies may be useful mainly as a concomitant one. It would be very interesting to develop mathematical models of these phenomena in order to guide the use of these therapies, specially after the recent results clinical trials 
investigating the combination of antiangiogenic therapies with the standard of care that have shown no impact on overall survival \citep{anti1,anti2}, though the progression free survival seems to benefit from the use of these drugs.

\par

It is important to remark that stochastic fluctuations in the therapy might affect the tumor dynamics.
These changes can be due to stochastic variations of the drug pharmaco-kinetics as shown by \citep{d'Onofrio2010}.
However, in this paper we have assumed that drug administration is such that the drugs achieve their therapeutical regimes that is preserved during treatment and constant in each simulation. 

\par

Other results from our simulations are that the different treatment modalities proposed suggest a more substantial effect on larger initial tumors that on smaller ones.  Thus, the combination of antithrombotics and antioxidants might be more effective for patients having only a biopsy or subtotal resections than for those in which the tumor could be macroscopically resected. In fact, for non resected tumors, a combination of antithrombotics and antioxidants could be more effective than radiotherapy alone probably due to the poor oxygenation and the presence of necrosic and hypoxic areas within the tumor. In addition, this combined therapy could even be more favorable for those patients with high risk of vessel coagulation.

\par

In conclusion, we have developed a biocomputational model of glioblastoma multiforme  that incorporates the spatio-temporal interplay among two glioma cell phenotypes corresponding to oxygenated and hypoxic cells, a necrotic core and the local vasculature whose response evolves with the tumor progression. Our  \emph{in silico} approach reveals that the different therapeutic modalities which combine antithrombotics and antioxidants would improve vessel functionality, avoiding vaso-occlusions and reducing oxidative stress and the subsequent hypoxic response. This combined treatment would reduce glioma cell invasion and sensitize glioblastoma to conventional radiotherapy, hopefully increasing patients survival.

\section*{Acknowledgements} This work has been partially supported by grants MTM2012-31073 (Ministerio de Econom\'{\i}a y Competitividad, Spain) and the James S. McDonnell Foundation (USA) through the 21st Century Science Initiative in Mathematical \& Complex Systems Approaches for Brain Cancer-Pilot Award  220020351.

\end{document}